\documentclass[letter,twocolumn]{jpsj2} 
\usepackage{latexsym}
\usepackage{bm}
\usepackage{graphicx}
\newcommand{\ket}[1]{|n(#1)\rangle}
\newcommand{\braket}[2]{\langle n(#1)|n(#2)\rangle}
\newcommand{\braOket}[3]{\langle n(#1)|#2|n(#3)\rangle}
\newcommand{\hatmu}{\hat\mu}

\newcommand{\rmB}{{\rm B}}

\newcommand{\U}{\mathop{\rm {}U}}

\title{Chern Numbers in Discretized Brillouin Zone:\\
Efficient Method of Computing (Spin) Hall Conductances}

\author{Takahiro Fukui,$^1$ Yasuhiro Hatsugai$^2$ and Hiroshi Suzuki$^3$}

\inst{$^1$Department of Mathematical Sciences, Ibaraki University, Mito
310-8512, Japan\\
$^2$ Department of Applied Physics, University of Tokyo, 7-3-1 Hongo, Bunkyo-ku,
Tokyo 113-8656, Japan\\
$^3$ Institute of Applied Beam Science, Ibaraki University, Mito 310-8512,
Japan}

\recdate{\today}

\abst{We present a manifestly gauge-invariant description of Chern numbers
associated with the Berry connection defined on a discretized Brillouin zone.
It provides an efficient method of computing (spin) Hall conductances without
specifying gauge-fixing conditions. We demonstrate that it correctly
reproduces quantized Hall conductances even on a coarsely discretized Brillouin
zone. A gauge-dependent integer-valued field, which plays a key role in the
formulation, is evaluated in several gauges. An extension to the non-Abelian
Berry connection is also given.}

\kword{Chern number, Berry connection, quantum Hall effect, lattice gauge
theory}

\begin{document}
\maketitle

Topological phase transitions have been of considerable interest in recent
condensed matter physics.\cite{Wen89,Hat04,Hat05} In lower dimensions, topological
quantum numbers are known to play a crucial role in characterizing various
phase transitions. A typical example is the integer quantum Hall transition,
where quantized Hall conductances are given by Chern numbers associated with
the Berry connection.~\cite{TKNN,Koh85,Ber84,Sim83} Its extension to the case
of spin currents is also attracting much  current interest.\cite{smf,mh,spin,Hal04} 
These topological quantum numbers present a chance to characterize quantum liquids
without using conventional symmetry breaking~\cite{Hat04,Hat05}.

Generically, the Chern numbers can be defined for quantum states with two
periodic parameters. As shown below, they are given by an integral of
fictitious magnetic fields (field strengths of the Berry connection) over
two-dimensional compact surfaces such as the Brillouin zone. In practical
numerical calculations, however, we can 
diagonalize Hamiltonians  only on a set of
discrete points chosen appropriately within the surfaces.
It is thus crucial to develop an efficient method of
calculating the Chern numbers using restricted data of wave functions 
given only on such discrete points.  
In these calculations, a phase ambiguity
of the wave function causes a gauge ambiguity for the Berry connection.
Therefore, one must be careful if gauge-dependent quantities are used. 

In this letter, we propose an efficient method of calculating the Chern numbers
on a discretized Brillouin zone. This is an application of a geometrical
formulation of topological charges in lattice gauge
theory.\cite{Luscher:1981zq,Phillips:1986us,Phillips:1986qd,PhiSto90,Fujiwara:2000wn}
We show that the Chern numbers thus obtained are {\it manifestly gauge-invariant\/} 
and {\it integer-valued\/} even for a discretized Brillouin zone. 
This implies that 
one can compute the Chern numbers using wave functions in
{\it any gauge\/} or {\it without specifying gauge fixing-conditions}. For
the purpose of demonstration, 
we apply our method to a simple model describing the integer
Hall system. We find that even for coarsely discretized Brillouin zones, the
method reproduces correct Chern numbers known so far. Our method can be
useful in a practical computation for more complicated systems with a
topological order for which a number of data points of the wave functions
cannot easily be increased.

To be specific, we focus on the Chern numbers in the quantum Hall effect. An
extension to different topological ordered states is straightforward. The spin
Hall conductances, for example, can be treated in a similar way. We consider a
two-dimensional system in which the Brillouin zone is defined by
$0\le k_\mu<2\pi/q_\mu$ ($\mu=1$, 2 with some integers~$q_\mu$). Since the
Hamiltonian~$H(k)$ is periodic in both directions,
$H(k_1,k_2)=H(k_1+2\pi/q_1,k_2)=H(k_1,k_2+2\pi/q_2)$, the (magnetic) Brillouin
zone can be regarded as a two-dimensional torus~$T^2$. When the Fermi energy
lies in a gap, the Hall conductance is given by
$\sigma_{xy}=-(e^2/h)\sum_nc_n$, where $c_n$ denotes the Chern number of the
$n$th Bloch band, and the sum over $n$ is restricted to the bands below the
Fermi energy.\cite{TKNN,Koh85} The Chern number assigned to the $n$th band is
defined by
\begin{equation}
c_n=\frac{1}{2\pi i}\int_{T^2}d^2k\,F_{12}(k),
\label{CheNumCon}
\end{equation}
where the Berry connection~$A_\mu(k)$ ($\mu=1$, 2) and the associated field
strength~$F_{12}(k)$ are given by\cite{TKNN,Ber84,Sim83}
\begin{eqnarray}
   &&A_\mu(k)=\braOket{k}{\partial_\mu}{k},
\nonumber\\
   &&F_{12}(k)=\partial_1A_2(k)-\partial_2A_1(k),
\label{cont-f}
\end{eqnarray}
with $\ket{k}$ being a normalized wave function of the $n$th Bloch band
such that $H(k)\ket{k}=E_n(k)\ket{k}$. In the above expressions, the
derivative~$\partial_\mu$ stands for~$\partial/\partial k_\mu$. We assume that
there is no degeneracy for the $n$th state.\cite{Hat04,Hat05} The phase of the wave
function is not yet determined here; that is, $\ket{k}$ is defined on~$T^2$
only up to its phase.

If the gauge potential~$A_\mu(k)$ is globally well defined over the continuum
Brillouin zone~$T^2$, the Chern number~(\ref{CheNumCon}) vanishes because the
torus has no boundary: It can be nonzero 
only when the gauge potential cannot be defined as a
global function over~$T^2$. In this
case, one covers $T^2$ by several coordinate patches and then, within each
patch, one can take a gauge (that is, a phase convention for the wave
functions) such that the gauge potential is a smooth and well defined 
function. In an overlap between two patches, gauge potentials defined on each
patch are related by a $\U(1)$ gauge transformation:
\begin{equation}
   \ket{k}\to e^{-i\lambda(k)}\ket{k},\quad A_\mu(k)\to A_\mu(k)
-i\partial _\mu\lambda(k).
\label{GauTra}
\end{equation}
The Chern
number~(\ref{CheNumCon}) is then given by a sum of the winding number of the
$\U(1)$ gauge transformation along a boundary of a patch. As a consequence, the
Chern number is an {\it integer}.

The above discussion is for the continuum Brillouin zone. Now suppose that we
have data of wave functions only on discrete points within the
Brillouin zone, as in actual numerical computations. A straightforward approach
for computing the Chern number~(\ref{CheNumCon}) would be to replace all the
derivatives by discrete differences and the integral by a summation. Namely,
one approximates the connection $A_\mu(k)dk_\mu$ by
\begin{equation}
   A_\mu(k)\delta k_\mu=\langle n(k)|\delta_\mu|n(k)\rangle ,
\label{three}
\end{equation}
where $\delta_\mu$ is an infinitesimal difference operator defined by 
$\delta_\mu f(k)=f(k+\hat{\delta k_\mu})-f(k)$ with $\hat{\delta k_\mu}$ being
an infinitesimal displacement vector in the direction~$\mu$ (its magnitude
is~$|\delta k_\mu|$).
Note that, to evaluate the difference, one must
fix a local gauge with which the state $|n(k) \rangle $ is
smoothly differentiable near $k$.
Under this local gauge, 
the field strength in the continuum is also approximated
by
\begin{equation}
   F_{12}(k)\delta k_1\delta k_2=[\delta_1A_2(k)-\delta_2A_1(k)]
   \delta k_1\delta k_2.
\label{four}
\end{equation}
Summing  this $F_{12}(k)\delta k_{1}\delta k_{2}$ then gives
the Chern number $c_n$ in the limit $|\delta k_{\mu} |\to 0$.
However, this direct procedure can be costly in taking the limit
if the Hamiltonian concerned is  complicated.

Here, we propose an alternative approach. Let us denote lattice points~$k_\ell$
($\ell=1$, \dots, $N_1N_2$) on the discrete Brillouin zone as
\begin{equation}
   k_\ell=(k_{j_1},k_{j_2}),\quad
   k_{j_\mu}=\frac{2\pi j_\mu}{q_\mu N_\mu},\quad
   (j_\mu=0,\ldots,N_\mu-1).
\end{equation}
We assume that the state $\ket{k_\ell}$ is periodic on the lattice,
$\ket{k_\ell+N_\mu\hatmu}=\ket{k_\ell}$, where $\hatmu$ is a vector in the
direction~$\mu$ with the magnitude~$2\pi/(q_\mu N_\mu)$. Below, we
set $N_\mu=q_\nu N_\rmB$~($\mu\ne\nu$) so that the unit plaquette is a square
of the size~$2\pi/(q_1q_2N_\rmB)$.

We first define a $\U(1)$ link variable from the wave functions of the $n$th
band as
\begin{equation}
   U_\mu(k_\ell)\equiv
   \braket{k_\ell}{k_\ell+\hatmu}/{\cal N}_\mu(k_\ell),
\label{seven}
\end{equation}
where ${\cal N}_\mu(k_\ell)\equiv|\braket{k_\ell}{k_\ell+\hatmu}|$. The link
variables are well defined as long as ${\cal N}_\mu(k_\ell)\neq0$, which can
always be assumed to be the case (one can avoid a singularity 
${\cal N}_\mu(k_\ell)=0$ by an
infinitesimal shift of the lattice). From the link variable~(\ref{seven}), we
next define a lattice field strength by
\begin{eqnarray}
   &&\tilde F_{12}(k_\ell)\equiv
   \ln U_1(k_\ell)U_2(k_\ell+\hat1)U_1(k_\ell+\hat2)^{-1}U_2(k_\ell)^{-1},
\nonumber\\
   &&-\pi<\frac{1}{i}{\tilde F}_{12}(k_\ell)\leq\pi.
\label{FieStr}
\end{eqnarray}
Note that the field strength is defined within the principal branch of the
logarithm specified in eq. (\ref{FieStr}). 
It is obvious that this field strength is invariant under the gauge
transformation~(\ref{GauTra}). Finally, we define the Chern number on the
lattice which is associated to the $n$th band as
\begin{equation}
   \tilde c_n\equiv\frac{1}{2\pi i}\sum_\ell\tilde F_{12}(k_\ell).
\label{CheNum}
\end{equation}

First of all, we note that $\tilde c_n$ is manifestly {\it gauge-invariant\/}
under eq.~(\ref{GauTra}). This implies that we do not need to determine
which gauge is adopted; any choice of gauge gives an identical
number~$\tilde c_n$. Moreover, $\tilde c_n$ is {\it strictly an integer\/} for
arbitrary lattice spacings. To show this, we introduce a gauge potential
\begin{equation}
   \tilde A_\mu(k_\ell)=\ln U_\mu(k_\ell),\quad
   -\pi<\frac{1}{i}\tilde A_\mu(k_\ell)\leq\pi,
\label{GauPot}
\end{equation}
which is periodic on the lattice:
$\tilde A_\mu(k_\ell+N_\mu\hatmu)=\tilde A_\mu(k_\ell)$. Recalling
definition~(\ref{FieStr}), one finds
\begin{equation}
   \tilde F_{12}(k_\ell)
   =\Delta_1\tilde A_2(k_\ell)-\Delta_2\tilde A _1(k_\ell)
   +2\pi in_{12}(k_\ell),
\label{FandN}
\end{equation}
where $\Delta_\mu$ is the forward difference operator on the lattice,
$\Delta_\mu f(k_\ell)=f(k_\ell+\hatmu)-f(k_\ell)$, and $n_{12}(k_\ell)$ is an
{\it integer-valued\/} field, which is chosen such that
$(1/i){\tilde F}_{12}(k_\ell)$ takes a value within the principal branch.
By definition, $|n_{12}(k_\ell)|\le2$. From eqs.~(\ref{CheNum})
and~(\ref{FandN}), we have
\begin{equation}
  \tilde c_n=\sum_{\ell} n_{12}(k_\ell),
\label{CAsN}
\end{equation}
which shows that the lattice Chern number~$\tilde c_n$ is an integer.

The field strength on the lattice $\tilde F_{12}(k_\ell)$ in eq.~(\ref{FieStr})
reduces to the one in the continuum~$F_{12}(k)\delta k_1\delta k_2$ in the
limit~$N_\rmB\to\infty$, where $\delta k_\mu=2\pi/(q_1q_2 N_\rmB)$.
Generically, the continuum field strength~$F_{12}(k)$ has no singularity when
the $n$th band is well separated from the neighboring ones; that is, the
energy gaps between them do not close,
\begin{equation}
   |E_n(k)-E_{n\pm1}(k)|\neq 0,
\label{gap}
\end{equation}
for any value of $k\in T^2$. This is the {\it gap-opening condition}.\cite{Hat04,Hat05}
 One can
expect, in general, that the problem is regular if the above gap-opening
condition is satisfied. Then, the lattice field strength~$\tilde F_{12}$ will
be small enough for a sufficiently large~$N_\rmB$ and the lattice Chern number
will approach the one in the continuum $\tilde c_n\to c_n$ in the
$N_\rmB\to\infty$ limit. Since both $\tilde c_n$ and~$c_n$ are integers, we
have $\tilde c_n=c_n$ for $N_\rmB>N_\rmB^c$. The critical mesh size~$N_\rmB^c$
may be estimated by a breaking of the admissibility
condition~\cite{Luscher:1981zq,Phillips:1986us,Phillips:1986qd,PhiSto90,Fujiwara:2000wn}
\begin{equation}
   |F_{12}(k_\ell)|\delta k_1\delta k_2\approx
   |\tilde F_{12}(k_\ell)|
   <\pi\quad\hbox{for all $k_\ell$}.
\label{fourteen}
\end{equation}
It is expected that this $N_\rmB^c$ is not very large for a standard generic
problem with the Chern number $c_n\approx{\cal O}(1)$. Since the area of the
Brillouin zone is~$4\pi^2/(q_1q_2)$, we can estimate the field
strength as $F_{12}(k_\ell)\approx ic_nq_1q_2/(2\pi)$. In this way, the
critical mesh size is given by
\begin{equation}
   N_\rmB^c\approx{\cal O}(\sqrt{2|c_n|/(q_1q_2)}).
\end{equation}
That is, we can expect that our method reproduces correct Chern numbers of the
continuum even for a coarsely discretized Brillouin zone. This is another
advantage of the present method.

As a function of $\U(1)$ link variables which satisfy the
admissibility~(\ref{fourteen}), the Chern number on the lattice~$\tilde c_n$ is
a constant function. To verify this, we note that a possible discontinuity
of~$\tilde c_n$ as a function of link variables $U_\mu(k_\ell)$ occurs only
when $|\tilde F_{12}(^\exists k_\ell)|=\pi$. Since $\tilde c_n$~is an integer
which cannot continuously change, $\tilde c_n$ remains the same as long as a
configuration is smoothly varied under the admissibility~(\ref{fourteen}). In
other words, under the admissibility, the space of $\U(1)$ link variables is
divided into disconnected sectors and the topological number~$\tilde c_n$ is
uniquely assigned to each sector. This is the basic idea behind the present
construction.~\cite{Luscher:1981zq} The Chern number~$\tilde c_n$ is, moreover,
a {\it unique\/} gauge-invariant topological integer which can be assigned to
admissible $\U(1)$~link variables.\cite{Luscher:1998du,footnote}

In the present context of the Berry connection, a gauge-invariant content of
link variables is completely governed by the $k$~dependence of the
Hamiltonian~$H(k)$. Each of the topological ordered states with a nontrivial Chern
number corresponds to the above nontrivial topological sector specified by the
admissibility. It is characterized by the lattice Chern number~$\tilde c_n$.
In the continuum, on the other hand, 
the topological stability of the Chern number is assured
by the gap-opening condition~(\ref{gap}).
 The topological quantum phase
transitions are thus characterized by the gap closing. Namely, nontrivial
topological sectors of the continuum, each of which is a topological ordered
state, are separated by the gaps.
Correspondingly the Chern number of the total bands, which is described by the
non-Abelian Chern number, vanishes.\cite{Hat04,Hat05}
At the critical point at which the gap-opening condition breaks down, 
the field strength $F_{12}(k)$ becomes singular at the gap-closing momentum. 
From a correspondence to the lattice case, we
conclude that the admissibility condition cannot be satisfied by any finite
$N_\rmB$ at that critical point.

Our method can be extended to the case of the non-Abelian Berry connection
${\cal A}=\psi^\dagger d\psi$, which is an $M\times M$ matrix-valued one-form
associated with a multiplet
$\psi=(|n_1\rangle,\ldots,|n_M\rangle)$.\cite{Hat04,Hat05,wz} 
The associated Chern number is defined by
$c_\psi=\int_S\mathop{\rm tr}d{\cal A}/(2\pi i)$, an integral over a
two-dimensional surface~$S$ with a generic (relaxed) gap-opening condition;
$E_n(k)\neq E_{n'}(k)$ for all $k$, where $n\in I$ and $n'\notin I$ for
$I=\{n_1,\ldots,n_M\}$.\cite{Hat04,Hat05}
 It turns out that the present lattice prescription is
valid if one substitutes the $\U(1)$ link variable by
\begin{equation} 
   U_\mu(k_\ell)=\frac{1}{{\cal N}_\mu(k_\ell)}
   \det\psi^\dagger(k_\ell)\psi(k_\ell+\hatmu)
\label{NonAbe}
\end{equation}
with the normalization constant
${\cal N}_\mu(k_\ell)\equiv|\det\psi^\dagger(k_\ell)\psi(k_\ell+\hatmu)|$.
We define the associated field strength and the Chern number on the
lattice~$\tilde c_\psi$ by the same expressions as those for the Abelian case,
eqs.~(\ref{FieStr}) and~(\ref{CheNum}). This $\tilde c_\psi$ shares  the
features of the Chern number in the Abelian case~$\tilde c_n$. For regular
problems, we have $\tilde c_\psi=c_\psi$ for a sufficiently fine
discretization~$N_\rmB>N_\rmB^c$.

Having observed desired properties of our definition of the lattice Chern number,
we now demonstrate how it works in a definite model. We consider the
Hamiltonian for spinless fermions in an external magnetic field:
$H=-t\sum_{\langle i,j\rangle}c_i^\dagger e^{i\theta_{i,j}}c_j$, where
the flux per plaquette on the coordinate
lattice~$\phi=\sum_\Box\theta_{i,j}/(2\pi)$ is~$p/q$.\cite{TKNN}
 For mutually prime
integers $p$ and~$q$, the spectrum splits into $q$~subbands.
 In the Landau
gauge in the $x$-direction, the Hamiltonian in the $k$-space is given by
$H_{ij}(k)=-2t\delta_{ij}\cos(k_y-2\pi\phi j)-t(\delta_{i+1,j}+\delta_{i,j+1})
-t\delta_{i+q-1,j}e^{-iqk_x}-t\delta_{i,j+q-1}e^{iqk_x}$,
($i$, $j=1$, \dots, $q$) with $q_1=q$ and $q_2=1$. Bellow, we will
present some results of applying our method to the middle subband of the $\phi=1/3$
(that is, $q=3$) system.
\begin{figure}[htb]
\begin{center}
\includegraphics[width=0.99\linewidth]{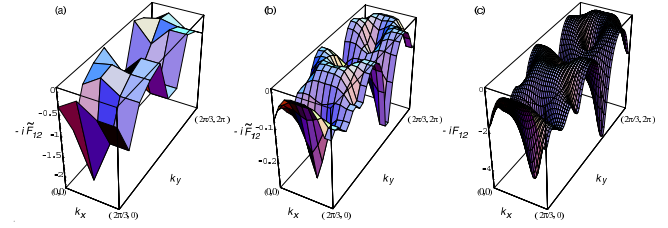}
\end{center}
\caption{(a) Field strength $-i{\tilde F}_{12}(k)$ of middle band
for $\phi=1/3$ system in $3\times9$ lattice Brillouin zone ($N_\rmB=3$),
(b) the same result in $9\times 27$ lattice ($N_\rmB=9$), and
(c) field strength $-i{F}_{12}(k)$ of the same band in the continuum
which is approximated by $\text{Im}\, ( \delta_{1}A_2-\delta_{2}A_1)  $ for 
$|\delta k_1|=|\delta k_2 |=2 \pi/90 $.}
\label{f:FieStr}
\end{figure}

In Fig.~\ref{f:FieStr}(a), we show the lattice field
strength~$\tilde F_{12}(k)$ in eq.~(\ref{FieStr}). The (magnetic) Brillouin zone
in the Landau gauge $[0,2\pi/3)\times[0,2\pi)$ is discretized by $3\times 9$
meshes. Note that the asymmetry of the Brillouin zone is simply due to the
gauge choice; there is no $x$-$y$ anisotropy in the present problem. The sum of
${\tilde F}_{12}(k)$ over the mesh points gives $\tilde c_n=-2$, which
coincides with the known result for the present case. The same calculation but
with $9\times 27$ meshes is shown in Fig.~\ref{f:FieStr}(b). It indeed gives
the identical Chern number~$\tilde c_n=-2$. Figure~\ref{f:FieStr}(c) shows
the field strength $F_{12}$ of the continuum. As expected, it is regular and of
the order of unity. Comparing these figures, one can see that the field
strength of the lattice system~$\tilde F_{12}$ converges to the one in the
continuum~$F_{12}$ up to a proportionality constant. Since the problem is
regular, the field strength of the lattice system~$\tilde F_{12}$ decreases as
$N_\rmB$~increases. The admissibility is safely satisfied in
Fig.~\ref{f:FieStr}(b) ($N_\rmB=9$), but $N_\rmB=3$ is close to the
critical $N_\rmB^c$ (Note the scales in the figures.)

It should be stressed again that the above lattice calculations can be
performed in {\it any gauge}. We do {\it not\/} need  specific gauge-fixing to
make the gauge connection smooth. An {\it arbitrary\/} gauge (e.g., a phase
choice of eigenvectors given by a numerical library) can be adopted to
compute the Chern number. 

As we have discussed, the lattice Chern number~$\tilde c_n$ is closely related
to the integer field~$n_{12}(k_\ell)$ in eq.~(\ref{FandN}). To illustrate this
point explicitly, we next plot the field~$n_{12}(k_\ell)$. Since we must
specify the gauge to do so ($n_{12}(k_\ell)$ itself is {\it not\/} 
gauge-invariant), we briefly describe the method of 
gauge-fixing adopted here.\cite{Hat04,Hat05}
One first selects an arbitrary state $|\phi\rangle$ which is globally
well defined over the whole Brillouin zone. Then the gauge can be specified by
$|n^\phi\rangle=P_n|\phi\rangle/N^\phi=
|n\rangle\cdot\langle n|\phi\rangle/N^\phi$, where $P_n=|n\rangle\langle n|$ is
a gauge-invariant projection and $N^\phi=|\langle\phi|n\rangle|$ 
is a gauge-invariant normalization which ensures $\langle n^\phi|n^\phi\rangle=1$.
A typical example of~$|\phi\rangle $ is a constant state, but it can be a
varying state as well.

\begin{figure}[htb]
\begin{center}
\includegraphics[width=0.999\linewidth]{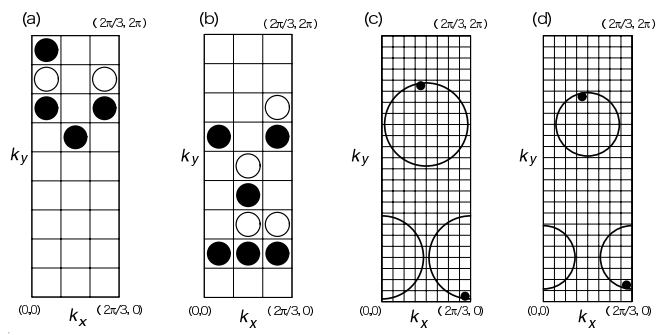}
\end{center}
\caption{Configuration of integer field~$n_{12}(k_\ell)$ in
gauge specified by state~$|\phi\rangle$ (see text) over discretized
Brillouin zones.
(a) $N_\rmB=3$, $|\phi\rangle=|\phi_{g_1}\rangle$
(b) $N_\rmB=3$, $|\phi\rangle=|\phi_{g_2}\rangle$
(c) $N_\rmB=8$ and $|\phi\rangle=|\phi(R_{\pi/3.2})\rangle$, and
(d) $N_\rmB=8$ and $|\phi\rangle=|\phi(R_{\pi/4.2})\rangle$. 
Black (white) circles denote $n_{12}=-1$ (1).}
\label{f:NFie}
\end{figure}

The integer fields $n_{12}(k_\ell)$ in several different gauges are depicted in
Fig.~\ref{f:NFie}, where the black and white circles denote $n_{12}=-1$
and~$1$, respectively, whereas a blank implies~$n_{12}=0$. 
It is clear that any of
them gives the correct Chern number~$\tilde c_n=-2$, that is, the number of black
circles minus that of white ones is always two. The field~$n_{12}(k_\ell)$ is
gauge-dependent, but their sum is gauge-invariant. The figures clearly
show the gauge-independence of the lattice Chern number.
In Figs.~\ref{f:NFie}(a) and \ref{f:NFie}(b), we used the global gauges specified by the
states $|\phi_{g_1}\rangle =e^{iq(k_x+k_y)}(1,-1,0)^T$ and 
$|\phi_{g_2}\rangle=e^{iq(k_x+k_y)}(1,1,0)^T$, respectively. The term ``global''
means that the gauge-fixing condition
$|\langle\phi_{g_i}(k_\ell)|n(k_\ell)\rangle|\neq0$ is satisfied at all lattice
points~$k_\ell$.

\begin{figure}[h]
\begin{center}
\includegraphics[width=0.99\linewidth]{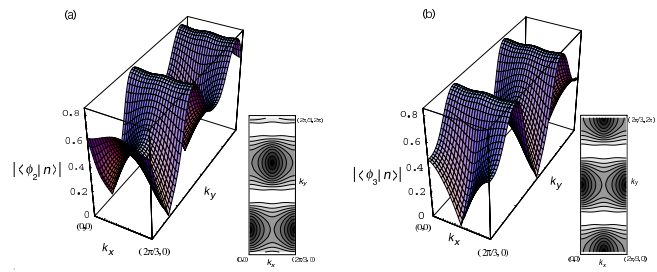}
\end{center}
\caption{Amplitude of overlap $|\langle\phi_i|n\rangle|$ between
wave function $|n\rangle$ and trial state $|\phi_i\rangle$:
(a) for trial state $\phi_2=(0,1,0)^T$, and
(b) for trial state $\phi_3=(0,0,1)^T$.}
\label{f:psi}
\end{figure}

The meaning of the field~$n_{12}(k_\ell)$ and the relationship between the
present lattice formulation and the continuum one become much clearer 
by adopting a ``patchwork gauge''.
To be specific, 
let us take a gauge convention specified by $|\phi_2\rangle=(0,1,0)^T$, 
but another gauge specified by $|\phi_3\rangle=(0,0,1)^T$
in some regions where the former is ill-defined.
We show in Fig.~\ref{f:psi} the amplitude of 
the overlap between the wave function and trial states
$|\phi_2\rangle$ and $|\phi_3\rangle$. 
Figure~3(a) shows that the gauge specified by~$|\phi_2\rangle$ becomes
ill-defined at two points near $k^1=(2\pi/3,\pi/3)$ and $k^2=(\pi/3,4\pi/3)$.
Therefore, we first define, around these points, circular
regions~$R_r=\{k\mid|k-k^1|<r\}\cup\{k\mid|k-k^2|<r\}$ 
with an appropriate radius~$r$, and we next check in Fig. \ref{f:psi}(b) 
that we can indeed 
take the second gauge specified by~$|\phi_3\rangle$ safely in~$R_r$. 
This patchwork gauge
choice is referred to as the gauge specified by~$|\phi(R_r)\rangle$. With this
gauge, the wave functions $|n^\phi(k_\ell)\rangle$ and the corresponding gauge
potential $\tilde A_\mu^\phi(k_\ell)$ are smooth if lattice points are
sufficiently fine.
This implies that the integer field $n_{12}(k_\ell)$ is vanishing
within each region. Nonzero values of~$n_{12}(k_\ell)$ are only allowed at
plaquettes existing at the boundary of the regions.

The field $n_{12}(k_\ell)$ computed with the above local gauge is shown in
Figs.~\ref{f:NFie}(c) and \ref{f:NFie}(d) for two different radiuses~$r$. The figures
clearly show that the integer field~$n_{12}(k_\ell)$ indeed acquires nonzero values
only at boundaries of separated regions. The lattice field $n_{12}(k_\ell)$
carries information corresponding to the winding number of the gauge transformation along the boundary of a patch in the continuum.

In this letter, we presented our method as an efficient technique for calculating
the Chern numbers in an infinite system on the basis of a discretized Brillouin
zone. In finite systems, however, the Brillouin zones are discrete from the
onset. Therefore, the present method will also be useful for revealing topological
orders of {\it finite systems\/} with possible many-body
interactions~\cite{Hat04,Hat05}.

The authors would like to thank Takanori Fujiwara for valuable discussions.
This work was supported in part by Grant-in-Aid's for Scientific Research from
JSPS.


\begin{thebibliography}{99}
\bibitem{footnote}
For higher dimensional lattices, this uniqueness of lattice Chern numbers holds
if the admissibility is modified to $|\tilde F_{\mu\nu}(k_\ell)|<\epsilon$ with
a certain constant~$\epsilon\leq\pi/3$.
\end{thebibliography}
\end{document}